\documentstyle[12pt]{article} 
\begin{document} 
\begin{center}
\vspace*{1.0cm} 
{\Large\bf An Extension of the Characteristic Angle Method
to the Easy-plane Spin-$\frac{3}{2}$ Ferromagnet} \\ \vspace*{1.0cm}
  Lei Zhou and Ning Xie\\ \vspace*{0.2cm} {\it Department of Physics,
Fudan University, Shanghai 200433, P. R.  China}\\ \vspace*{0.6cm} Ruibao
Tao\\ \vspace*{0.2cm} {\it Center for Theoretical Physics, Chinese Center
of Advanced Science and Technology (World Laboratory) ,
 P. O. Box 8730,
 Beijing 100080, China }\\ {\it and Department of Physics, Fudan
University,
 Shanghai 200433, P. R. China}\\ \end{center}

\vspace*{0.5cm}


\vspace*{0.5cm}

\centerline{\bf\large Abstract}
\vspace*{1.0cm}
The Characteristic Angle (CA) method [Lei Zhou and Ruibao Tao, J. Phys. A
{\bf 27} 5599]
developed previously for the easy-plane spin-1 magnetic systems
has been successfully extended to the spin-$\frac{3}{2}$ case. A compact
form of the CA spin-$\frac{3}{2}$ operator transformation is given,
then the ground state energy, the magnon dispersion relation and the
spontaneous magnetization are discussed for an easy-plane
spin-$\frac{3}{2}$ ferromagnet by using the CA method.
Comparisons with the old theoretical methods
are made in the end.

\newpage
\section{Introduction}
``Easy-plane" single-ion anisotropy ($D(S^x_i)^2, D>0$) is very
popular in magnetic systems with spin greater than one half,
and such systems have been attracting much attention for
years [1-11]. On theoretical side,
the spin-wave excitations in such systems are very
difficult to handle caused by the off-diagonal terms introduced by the
``easy-plane" single-ion anisotropy term. Actually, many straightforward
methods which are good to deal with the ``easy-axis" single-ion anisotropy
($D(S^x_i)^2, D<0$) are usually found to be invalid for the
``easy-plane"
magnetic systems no matter how weak the anisotropy parameter $D$ will
be [4,11-13]. For example, applying the Holstein-Primakoff (H-P)
transformation [12] naively to study the magnon excitations in the
easy-plane magnetic systems must lead to an imaginary value for
the energy of the ``k=0" mode [4,11], and introducing a
vector rotation of the spins in the XZ plane to optimize the spontaneous
magnetized direction is also helpless [13] -
the excitation energy in a harmonic approximation
under such an approach will either be negative or be imaginary.
In fact, caused by the off-diagonal terms
($D((S^+_i)^2 + (S^-_i)^2)$) in the Hamiltonian introduced by the
single-ion anisotropic terms, the proper eigenstate of the Hamiltonian
must be a mixture of the single-site spin states $|n\rangle$ with
$|n+2\rangle$ and $|n-2\rangle$ [3,4].
Such a spin-states mixing effect is very significant in the present
``easy-plane" case, as the result, an ordinary method which fails to
consider such an effect must be invalid for such systems.
The matching of matrix elements (MME) method is one possible
way to take this effect into account perturbatively [3,4],
and many applications of this method have been implemented
in a first order approximation of the anisotropy term [3-7].
Some numerical methods had also been developed for the easy-plane spin-1
ferromagnetic systems where the authors had considered the spin-states
mixing effect [9-10]. Recently, a new method - the characteristic angle
(CA) method was developed successfully for the easy-plane
spin-1 magnetic systems [1], in which the spin operators are transformed
to a new set of quasi-spin operators which have
taken account of the spin-states mixing effect by a variation
parameter - the characteristic angle (CA).
The ground state properties of an easy-plane spin-1 ferromagnet,
such as the ground state energy, the spontaneous magnetization
and the magnon dispersion relation, are more reasonable in the CA
approach than those in the MME method [1]. The induced magnetization of the
easy-plane spin-1 ferromagnet has also been studied based on such an
approach [11].

In this paper, we would like to extend the CA method to the easy-plane
spin-$\frac{3}{2}$ magnetic systems. Although the main idea has
been proposed in Ref. 1, an extension of this
approach to larger spin case is still nontrivial because the extension
makes it possible to study the magnetic systems with single-ion anisotropy
in larger spin case. Actually, the extension is not very easy because
the number of the variation parameters will increase and the expression
of the CA transformation will become more complicated than that in spin-1
case. This paper is organized as follows. After diagonalizing the
single-site Hamiltonian part, we derive the CA spin operator transformation
for $S=\frac{3}{2}$ case successfully in Section 2, then we present
the main physical results for an easy-plane spin-$\frac{3}{2}$
ferromagnet using the CA method in Section 3, and the comparisons
with the MME method and other methods are made thereafter.
Finally, conclusions are outlined in Section 4.

\section{Characteristic Angle transformation for $S=\frac{3}{2}$}
The Hamiltonian of an easy-plane spin-$\frac{3}{2}$ ferromagnet
can be given as
\begin{eqnarray}
 H=-J \sum_{ (i,j) } {\bf  S}_i \bullet {\bf  S}_j
        +D \sum_{i} ( S_{i}^{x} )^{2} - h\sum_i S^z_i,
\end{eqnarray}
where $(i,j)$ means summation restricted on the nearest-neighbor pairs.
The spin vectors are forced into the YZ plane by the anisotropy term
($D>0$), so that the spontaneous magnetized direction must be
in the YZ plane. Without loosing any generality,
the Z direction is chosen to be the spontaneous magnetized
direction and an external magnetic field $h$ will be applied along the Z
direction which will be set to zero in the end.

Following Ref. 1, let us start from diagonalizing
the single-site part of the Hamiltonian.
 
The matrix form of the single-site Hamiltonian $D(S^{x})^{2}-h S^{z}$
in the $S^z$ representation is found to be
\begin{equation}
    {\cal A}=
    \left( \begin{array}{cccc}
\frac{3}{4}D-\frac{3}{2}h & 0 & \frac{\sqrt{3}}{2}D & 0\\
0 & \frac{7}{4}D-\frac{1}{2}h & 0 & \frac{\sqrt{3}}{2}D\\
\frac{\sqrt{3}}{2}D & 0 & \frac{7}{4}D+\frac{1}{2}h & 0\\
0 & \frac{\sqrt{3}}{2}D & 0 & \frac{3}{4}D+\frac{3}{2}h\\
  \end{array} \right)
\end{equation}
where the matrix elements are defined by
\begin{math}
{\cal A}_{mn} = \langle m|D(S^{x})^{2}-h S^{z}|n\rangle
\end{math} ~
in which the bases $|m\rangle$ are the complete set of
eigenstates of operator $S^z$ with eigenvalues
$(\frac{3}{2},\frac{1}{2},-\frac{1}{2},-\frac{3}{2})$.

A new representation can be defined as follows, in which the complete
set of bases $|\tilde m\rangle$ are related to the old ones by the
following orthogonal transformation:
\begin{equation}
|\tilde m \rangle = \sum_n |n \rangle {\cal P}_{nm},
\end{equation}
\begin{equation}
    {\cal P}=
    \left( \begin{array}{cccc}
\cos\theta_1  & 0             & \sin\theta_1 & 0\\
0             & \cos\theta_2  &   0          & -\sin\theta_2\\
-\sin\theta_1 & 0             & \cos\theta_1 & 0\\
0             & \sin\theta_2  &   0          & \cos\theta_2\\
  \end{array} \right)\label{eq:pm}.
\end{equation}

Then, after the transformation, the matrix form of the operator
$D(S^{x})^{2}-hS^{z}$ in the new representation $|\tilde m\rangle$
will be
\begin{eqnarray}
\tilde {\cal A} = {\cal P}^{T} {\cal A} {\cal P}.
\end{eqnarray}

It is easy to check that only when
\begin{eqnarray}
\tan 2\theta_1 =\frac{\sqrt{3} D}{D+2h},\label{eq:tan1}\\
\tan 2\theta_2 =\frac{\sqrt{3} D}{D-2h},\label{eq:tan2}
\end{eqnarray}
the matrix $\tilde {\cal A}$ can be exactly diagonalized with the
following eigenvalues
\begin{equation}
\left\{
\begin{array}{ll}
\lambda_1&=-\frac{3}{2}D-\frac{1}{2}h-\sqrt{D^2+h^2+Dh}\\
\lambda_2&=-\frac{3}{2}D+\frac{1}{2}h+\sqrt{D^2+h^2-Dh}\\
\lambda_3&=-\frac{3}{2}D-\frac{1}{2}h+\sqrt{D^2+h^2+Dh}\\
\lambda_4&=-\frac{3}{2}D+\frac{1}{2}h-\sqrt{D^2+h^2-Dh}\\
\end{array}
\right.
\end{equation}

Since the orthogonal transformation ${\cal P}$ can exactly diagonalize the
single-site part of Hamiltonian, then if we apply
this transformation to the total Hamiltonian (1), we can expect to get
a reasonable representation of the easy-plane spin-$\frac{3}{2}$
ferromagnet after selecting appropriate values of the parameters
$\theta_1,\theta_2$ by the variation method.

However, for the sake of simplification in this paper, we will only
study the easy-plane spin-$\frac{3}{2}$ ferromagnet in zero external
magnetic field here (in another word, $h=0$), then the two variation
parameters $\theta_1$ and $\theta_2$ should be the same.
In fact, since the Z direction is not specified in Hamiltonian (1),
the state that all spins are at $|\frac{3}{2}\rangle$
and the state that all spins are at $|-\frac{3}{2}\rangle$ will
give the same physical results. So, from the definition of the
transformation $\cal P$ (Eq. (\ref{eq:pm})), we may easily
obtain $\theta_1 = \theta_2 = \theta$. Of cause, if an external magnetic
field $h$ is applied, the two parameters may not be the same and things
are somewhat complicated then.

In order to investigate how the transformation acts on the exchange
interactions, we should first study how the transformation acts on the
spin operators. The matrix form of the spin operator $S^+$ in the
old $S^z$ representation is
\begin{equation}
    (~~S^+~~)=
    \left( \begin{array}{cccc}
0             & \sqrt{3}      & 0            & 0\\
0             & 0             & 2            & 0\\
0             & 0             & 0            & \sqrt{3}\\
0             & 0             & 0            & 0\\
  \end{array} \right).
\end{equation}

After the orthogonal transformation, the matrix form of
operator $S^+$ in the new representation is found to be
\begin{equation}
    (~~\tilde S^+~~)= {\cal P}^{T} (~~S^+~~) {\cal P}=
    \left( \begin{array}{cccc}
0 & \sqrt{3}\cos 2\theta  & 0 & -\sqrt{3}\sin 2\theta\\
-\sin 2\theta  & 0 & \sin 2\theta  & 0\\
0 & \sqrt{3}\sin 2\theta  & 0 & \sqrt{3}\cos 2\theta\\
2\sin^2 \theta & 0 & -\sin 2\theta & 0\\
  \end{array} \right)\label{eq:sp1}.
\end{equation}

Now we may understand this transformation in another way: suppose
the complete bases $|m\rangle$ are maintained unchanged,
instead the spin operators are transformed to new ones by an
unitary transformation. Noticing some matrix identities
for spin-$\frac{3}{2}$ operators such as
\begin{equation}
    (~~\sin(\pi S^z)~~)=
    \left( \begin{array}{cccc}
-1            & 0             & 0            & 0\\
0             & 1             & 0            & 0\\
0             & 0             & -1           & 0\\
0             & 0             & 0            & 1\\
  \end{array} \right),
\end{equation}
and
\begin{equation}
    (~~ (S^+)^3~~ )=
    \left( \begin{array}{cccc}
0             & 0             & 0            & 6\\
0             & 0             & 0            & 0\\
0             & 0             & 0            & 0\\
0             & 0             & 0            & 0\\
  \end{array} \right),
\end{equation}
we can easily get the following spin-operator transformation:
\begin{eqnarray}
P^{\dag} S^+ P  =  \tilde S^+ &=&~~~\frac{1}{2}\cos2\theta
[1+\sin(\pi S^z)]S^+
+\frac{1}{4}\sin2\theta[1-\sin(\pi S^z)]S^+\nonumber\\
& &-\frac{\sqrt{3}}{3}\sin2\theta S^- [1+\sin(\pi S^z)]
+\frac{\sqrt{3}}{2}\sin2\theta S^- [1-\sin(\pi S^z)]\nonumber\\
& &-\frac{\sqrt{3}}{6}\sin2\theta (S^+)^3
+\frac{1}{3}\sin^2 \theta (S^-)^3,\label{eq:sp}\\
P^{\dag} S^- P  =  \tilde S^- &=&~~~\frac{1}{2}\cos2\theta
S^-[1+\sin(\pi S^z)]
+\frac{1}{4}\sin2\theta S^-[1-\sin(\pi S^z)]\nonumber\\
& &-\frac{\sqrt{3}}{3}\sin2\theta [1+\sin(\pi S^z)]S^+
+\frac{\sqrt{3}}{2}\sin2\theta[1-\sin(\pi S^z)]S^+\nonumber\\
& &-\frac{\sqrt{3}}{6}\sin2\theta (S^-)^3
+\frac{1}{3}\sin^2 \theta (S^+)^3,\\
P^{\dag} S^z P  =  \tilde S^z
&=&\frac{1}{2}[\tilde S^+, \tilde S^-]_- \label{eq:sz}.
\end{eqnarray}

Eqs. (\ref{eq:sp})-(\ref{eq:sz}) are just the Characteristic
Angle (CA) spin operator
transformation for the spin-$\frac{3}{2}$ case, and $\theta$ is the
Characteristic Angle (CA) which is actually a variation parameter
determined later by minimizing the ground state energy.
The transformed spin operators, so called CA spin operators
($\tilde S^{\pm},\tilde S^z$),
obey all the spin-$\frac{3}{2}$ operator's commutation rules.
The proof of the above CA transformation is straightforward.
One may only write out the matrix form  of the CA spin operator
$\tilde S^+$ (Eq. (\ref{eq:sp})) in the ordinary $|m\rangle$ representation,
then can find out it is the same as the matrix on the right side of
Eq. (\ref{eq:sp1}). This means the CA transformation for the spin operators
is actually equivalent to the orthogonal transformation for the spin space.
Furthermore, since an orthogonal transformation does not affect the
commutation relations for matrices, and since the matrix form of the CA
spin operator is related to the matrix form of the original
spin operator by an orthogonal transformation,
 the CA spin operator ($\tilde S^{\pm}, \tilde S^z$)
must obey all the spin-$\frac{3}{2}$ operators' commutation rules.

By the way, although it seems possible to obtain the CA transformations
for arbitrary spin case following the method described in
this section, however, a closed form of the CA transformation is not
easy to derive for larger spin case because more variation parameters will
be needed then. An extension of this method to the general case
is very difficult.

\section{Results and Comparisons}

In order to study the ground state properties and the low-lying
excitations for such a system, the H-P transformation is introduced for
the spin-$\frac{3}{2}$ operators:
\begin{eqnarray}
 S_i^{z}  &=& \frac{3}{2}-a_i^{+}a_i,\\
 S_i^{+}  &=&\sqrt{3-a_i^{+}a_i}~~a_i,\\
 S_i^{-}  &=& a_i^{+} \sqrt{3-a_i^+a_i}.
\end{eqnarray}
Applying the above H-P transformation, we can obtain
the Bose expansions for CA spin-$\frac{3}{2}$ operators.

However, there exist another straightforward method to
determine the first several order terms of the Bose
expansions for the CA spin operators. We will
consider $\tilde S^+$ as an example. After transforming
the Hilbert space of spin operator to that of the Bose operator, the
vacuum state of Bose operator $|0\rangle_B$ is defined to be the spin
state $|\frac{3}{2}\rangle_S$, and the excited Bose states
$|1\rangle_B$, $|2\rangle_B$, $|3\rangle_B$ are the corresponding spin
states:
$|\frac{1}{2}\rangle_S$,$|-\frac{1}{2}\rangle_S$,$|-\frac{3}{2}\rangle_S$,
respectively (The subscript B and S are used to denote the Bose state
and the spin state, respectively). Other Bose
states such as $|n\rangle_B, n>3$ are unphysical states.
In physical space, the matrix forms of some
lower order Bose expansion can be written out readily:
\begin{eqnarray}
    (~~ a~~ )=
    \left( \begin{array}{cccc}
0             & 1             & 0            & 0\\
0             & 0             & \sqrt{2}     & 0\\
0             & 0             & 0            & \sqrt{3}\\
0             & 0             & 0            & 0\\
  \end{array} \right),\\
    (~~ a^2~~ )=
    \left( \begin{array}{cccc}
0             & 0             & \sqrt{2}     & 0\\
0             & 0             & 0            & \sqrt{6}\\
0             & 0             & 0            & 0\\
0             & 0             & 0            & 0\\
  \end{array} \right),
\end{eqnarray}
etc. Then from the matrix form of the operator $\tilde S^+$
(Eq. (\ref{eq:sp1})), it is very easy to write out the first
order terms of CA spin operator $\tilde S^+$:
\begin{eqnarray}
\tilde S^+ = \sqrt{3}\cos2\theta~ a - \sin2\theta~ a^+ + \cdots
\label{eq:spb}
\end{eqnarray} 
The Bose expansion of $\tilde S^z$ can be similarly derived as:
\begin{eqnarray}
\tilde S^z = \frac{3}{2}-2\sin^2\theta - a^+a
+\frac{\sqrt{2}}{2} \sin2\theta (a^{+ 2} + a^2) + \cdots \label{eq:szb}
\end{eqnarray} 
Other terms can be obtained easily.

Applying the CA transformation to Hamiltonian (1), we get
\begin{eqnarray}
\tilde H =-J \sum_{ (i,j) } {\bf {\tilde S}}_i \bullet {\bf {\tilde S}}_j
        +D \sum_{i} ( {\tilde S}_{i}^{x} )^{2} \label{eq:ht}.
\end{eqnarray}
The transformed Hamiltonian $\tilde H$ must have the same eigenvalues
with the
original one since the transformation is an orthogonal one.

Then we can apply the H-P transformation to the Hamiltonian
Eq. (\ref{eq:ht}) keeping a harmonic approximation. We get
\begin{eqnarray}
\tilde H &=&U_0 + H_2 + \cdots\\
U_0      &=&[-JZ(\frac{3}{2}-2\sin^2\theta)^2 +
             (\frac{7D}{4}-D\cos^2\theta-\frac{\sqrt{3}}{2}D\sin2\theta)]
               N,\\
H_2      &=&J\sum_{(i,j)}\{(3-4\sin^2\theta)[a_i^+a_i
             -\frac{\sqrt{2}}{2}\sin2\theta((a^+_i)^2+(a_i)^2)]\nonumber\\
         & &~~~~~-[(3\cos^2 2\theta + \sin^2 2\theta) a^+_i a_j +
              \sqrt{3}\sin2\theta\cos2\theta(a_i^+a^+_j + a_ia_j)] \}\nonumber\\
         & &+D\sum_i \{\frac{\sqrt{2}}{2}
             (\frac{\sqrt{3}}{2}\cos2\theta-\frac{1}{2}\sin2\theta)
             [(a^+_i)^2 + (a_i)^2]\nonumber\\
         & &~~~~~~~+ (\sqrt{3}\sin2\theta + \cos2\theta)~~a^+_ia_i \}.
\end{eqnarray}

Taking a Fourier transformation to the momentum ${\bf k}$ space,
we get the following Hamiltonian:
\begin{eqnarray}
\tilde H &=& U_0 + \sum_{k} A_k a^+_k a_k
             + \sum_k B_k (a^+_ka^+_{-k} + a_ka_{-k}) +
             \cdots\label{eq:h2k}\\
A_k      &=& JZ(3-4\sin^2\theta)-JZ(3\cos^2 2\theta + \sin^2 2\theta)
                \gamma_k \nonumber\\
         & & + \sqrt{3}D\sin2\theta + D\cos2\theta,\\
B_k      &=& \frac{\sqrt{2}}{2}[-JZ(3-4\sin^2\theta)\sin2\theta +
                \frac{\sqrt{3}}{2}D\cos2\theta
                -\frac{1}{2}D\sin2\theta]\nonumber\\
         & &+ JZ\sqrt{3}\sin2\theta\cos2\theta\gamma_k,
\end{eqnarray}
where
\begin{eqnarray}
\gamma_k = \frac{1}{Z}\sum_{\delta} exp(i{\bf k}\cdot {\bf r}),
\end{eqnarray}  
in which the $\delta$ summation runs over the $Z$ nearest-neighbor
sites of a given site.

The approximate Hamiltonian Eq. (\ref{eq:h2k}) can be diagonalized
by a Bogolyubov transformation,
\begin{eqnarray}
\tilde H   &=& E_0 + \sum_{k}\epsilon_k \alpha^+_k \alpha_k + \cdots\\
E_0        &=& U_0 - \frac{1}{2}\sum_k A_k +
                \frac{1}{2}\sum_k\sqrt{A^2_k-4B^2_k},\label{eq:e0}\\
\epsilon_k &=& \sqrt{A^2_k - 4 B^2_k}\label{eq:ek}.
\end{eqnarray}
The spontaneous magnetization for the easy-plane spin-$\frac{3}{2}$
ferromagnet in the harmonic approximation can be derived as:
\begin{eqnarray}
M_0 = \langle 0 | \tilde S^z |0\rangle
    = 2-2\sin^2\theta-\frac{1}{N}\sum_k
        \frac{\displaystyle\frac{1}{2}A_k+\sqrt{2}\sin2\theta B_k}
          {\sqrt{A^2_k-4B^2_k}}\label{eq:m0}.
\end{eqnarray}

As we know, the value of variation parameter $\theta$ should be
determined by minimizing th ground
state energy $E_0$. An analytical expression of $\theta$ with respect to
the anisotropy parameter seems very
difficult to obtain, however, an expansion in $d=D/JZ$ can be determined
order by order. As a first step, suppose $d$ is very small so that
$\theta$ is also small, then by minimizing the ground state energy keeping
only the first order terms in $d$, we obtain the following equation:
\begin{eqnarray}
          \frac{d}{d \theta} E \simeq \frac{d}{d \theta} U_0
 &=     &     -2JZ\sin4\theta - (2JZ + D)\sin2\theta
           +\sqrt{3}D\cos2\theta\nonumber\\
 &\simeq& - 2JZ \cdot 4\theta - 2JZ \cdot 2\theta + \sqrt{3} D = 0,
\end{eqnarray}
then,
\begin{eqnarray}
\theta \simeq \frac{\sqrt{3}}{12} d.
\end{eqnarray}

Applying this solution into Eqs. (\ref{eq:spb})-(\ref{eq:szb}),
it is easy to obtain
\begin{eqnarray}
\tilde S^+ &=& \sqrt{3}a - \frac{\sqrt{3}}{6}d~a^+ \cdots\label{eq:spd}\\
\tilde S^- &=& \sqrt{3}a^+ - \frac{\sqrt{3}}{6}d~a \cdots\\
\tilde S^z &=& \frac{3}{2} - a^+ a + \frac{\sqrt{6}}{12}d(a^{+2}+ a^2)
                \cdots\label{eq:szd},
\end{eqnarray}
then from Eqs. (\ref{eq:e0})-(\ref{eq:ek}),
we get the ground state energy and the magnon
dispersion relation as following:
\begin{eqnarray}
E_0        &=&-JZN(\frac{15}{4}-\frac{1}{4}d) -\frac{1}{2}JZ\sum_k
                \sqrt{[3(1-\gamma_k)+ d]^2 -
                (d\gamma_k)^2},\label{eq:re0}\\
\epsilon_k &=& JZ\sqrt{[3(1-\gamma_k)+ d]^2 -
(d\gamma_k)^2}\label{eq:rek}.
\end{eqnarray}
It is easy to check that $\epsilon_k \longrightarrow 0$ when
$k \longrightarrow 0$.

Comparing the transformation Eqs. (\ref{eq:spd})-(\ref{eq:szd})
and the results Eqs. (\ref{eq:re0})-(\ref{eq:rek}) to those of
the MME method which are obtained in a first order
approximation of $d$ [4,6], we are interested to find that
they are exactly the same.

In the case of large anisotropy $d$, the first order approximation
is not enough, numerical calculation must be required. We have
studied a simple-cubic lattice in the present paper, the characteristic
angle $\theta$, the
ground state energy $E_0$ and the spontaneous magnetization $M$ have
been shown in figures 1-3 as functions of the anisotropy
parameter $D / JZ$. The most interesting thing is that the easy-plane
spin-$\frac{3}{2}$ ferromagnet does not exhibit a phase transition from
the ferromagnetic phase to the non-ferromagnetic one
as $D / JZ$ is enlarged (figure 3). This is quite different
from the spin-1 systems [1,9]. Actually, when $D \gg JZ$, the single-ion
term is dominant. In this case, one may find from
Eqs. (\ref{eq:tan1})-(\ref{eq:tan2}) that $\theta=\pi /6$.
According to Eq. (\ref{eq:szb}), we get: when $D \gg JZ$,
$M \longrightarrow 1$. However, in spin-1 systems, one may find that
when $D > D_c (< 4JZ) $, $M \longrightarrow 0$ [1,9].

\vspace*{1.0cm}

In the remaining part of this section, we will compare the CA method
with the MME method and other old methods in details.
The total Hamiltonian of an easy-plane spin-$\frac{3}{2}$ ferromagnet
can be divided into two parts:
\begin{eqnarray}
H      &=&-J\sum_{(i,j)}(S^z_iS^z_j + S^+_iS^-_j) + D\sum_i (S^x_i)^2
           = H_0 + H_{int} + const.\\
H_0    &=& -2JZS\sum_i S^z_i + D\sum_i (S^x_i)^2,\label{eq:hb}\\
H_{int}&=& -J\sum_{(i,j)}[S^+_iS^-_j + (S^z_i - S)(S^z_j - S)].
\end{eqnarray}

The main question in such systems is to find a best
``starting point" to perform the spin waves calculation.
It is just because the ``starting point" are chosen unreasonably
that the methods mentioned in the introduction are invalid
for the easy-plane magnetic systems [4,11-13]. For example, the
``starting point" in H-P
transformation method is the ordinary ferromagnetic state
that all spins are aligned parallel to each other, however, this
``starting point"  does not consider the spin-states mixing effect
at all so that it must be unreasonable.

According to the MME method [4], the ``starting point" is
the ferromagnetic state in the diagonalized representation of
Hamiltonian $H_0$ (Eq. (\ref{eq:hb})).
Treating the anisotropy term $D(S^x_i)^2$
as a perturbation, the Hamiltonian $H_0$ can be diagonalized in terms of
$d=D/JZ$ order by order. Based on this, the spin-Bose transformation was
applied to the total Hamiltonian to study the magnetic properties
of the system. In such an approach, it should be noted
that the $H_{int}$ part of the Hamiltonian has
been neglected when determining the ``starting point", so
this approximation may be valid only when the anisotropy is very small.

However, we believe that the ``starting point" in the CA approach
will be more reasonable because the $H_{int}$ has also been taken
into account when determining the ``starting point".
As we can see in this paper, the CA transformation is applied to the
total Hamiltonian and $\theta$ is
determined by minimizing the total ground state energy $E_0$ (Eq.
(\ref{eq:e0})).
Actually, $H_{int}$ has been considered through its contribution
to $E_0$. Since $H_{int}$ will contribute the
second order terms in $d$ to the ground state energy $E_0$,
it is very easy to understand that the CA method gives
the same physical results as the MME method in the first order
approximation of $d$ which has been demonstrated in the present work.
However, for large $d$ case when higher order terms in $d$ should
be considered, the CA method may give more accurate results than the MME
method because the CA method has considered more contribution
than the MME method when choosing the ``starting point".

Of cause, the CA method can only deal with the
spin-1 and the spin-$\frac{3}{2}$ cases at present time, however,
the MME method can deal with arbitrary spin case perturbatively.

\section{Conclusions}
In this paper, we have successfully extended the characteristic angle
method to the easy-plane spin-$\frac{3}{2}$ magnetic systems.
Based on the diagonalization of the single-site part of the Hamiltonian,
a compact form of the CA transformation for the spin-$\frac{3}{2}$
operators has been derived, in which the spin-states mixing effect
is considered automatically by the variation parameters.
After applying the CA transformation and the H-P spin-Bose
transformation, a model Hamiltonian which has the same eigenvalues
with the original Hamiltonian is presented by the Bose operators.
Bogolyubov transformation is applied to diagonalize the Hamiltonian in
harmonic approximation, and the variation
parameter CA is determined by minimizing the ground state energy.
Analytical results of the magnetic properties are obtained in a first
order approximation of $d=D/JZ$ which are found to recover the MME method.
Numerical calculations are carried out in large $d$ case for
the easy-plane spin-$\frac{3}{2}$ ferromagnet, and detailed
comparisons with the MME method and other methods are made
in the end. We find that the CA method has considered more
contributions than the MME method and other methods for the
spin-1 and spin-$\frac{3}{2}$ easy-plane magnetic systems.

\vspace*{1.0cm}

\noindent {\bf Acknowledgments}\\

This research is supported by National Science Foundation of China and
the National Education Commission under the grant for training Ph.Ds.

\newpage
\vspace*{1.0cm}
\noindent{\bf Captions:}\\

\vspace*{1cm}

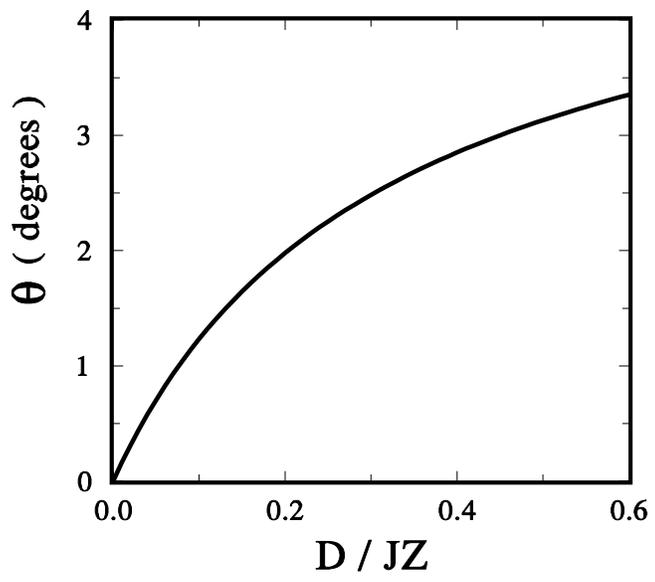
\noindent Figure 1:
\parbox[t]{14cm}{
Value of the characteristic angle as the function of the anisotropy
parameter.}

\vspace*{0.5cm}

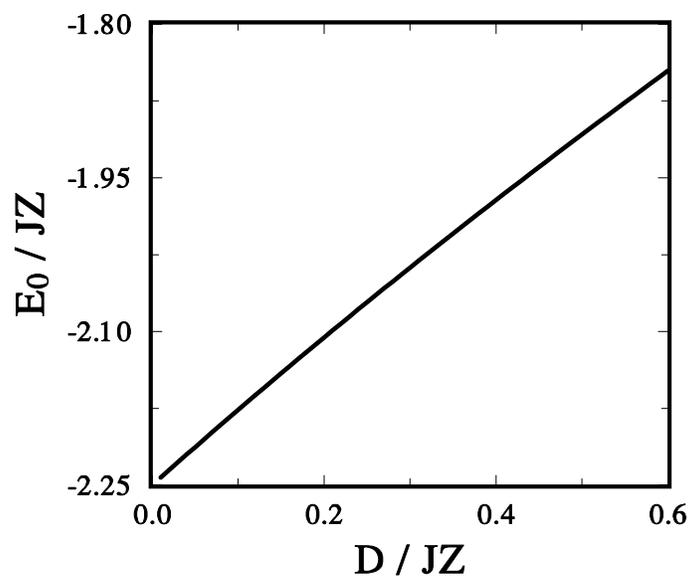
\noindent Figure 2:
\parbox[t]{14cm}{The ground state energy as the function of the
anisotropy parameter.}

\vspace*{0.5cm}

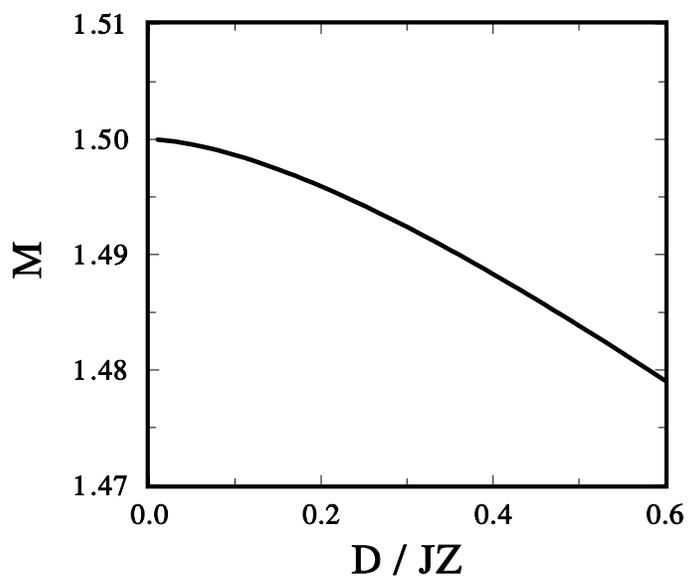
\noindent Figure 3:
\parbox[t]{14cm}{
The spontaneous magnetization as the function of the anisotropy
parameter.}


\newpage
\begin{thebibliography}{99}
\vspace*{1.0cm}
\bibitem{}    Lei Zhou and Ruibao Tao,
               J. Phys. A: Math and Gen. {\bf 27} (1994) 5599.
\bibitem{}  {\it Crystalline Electric Field and Structure Effect in
                f-Electron Systems},edited by Crow J E,
                Gruertin R P, and  Mihalisin T W (Plenum, New York, 1980)
\bibitem{}     Lindg\.{a}rd P-A and Danielsen O, J. Phys. C: Solid St.
              Phys. {\bf 7} (1974) 1523.
\bibitem{}     Lindg\.{a}rd P-A and Kowalska A, J. Phys. C: Solid St.
              Phys. {\bf 9} (1976) 2081.
\bibitem{}     Cooke J F and Lindg\.{a}rd P-A, Phys. Rev. {\bf B16}
                (1977) 408; J. Appl. Phys. {\bf 49} (1978) 2136.
\bibitem{}     U Balucani, M G Pini, A Rettori and V Tognetti
               J. Phys. C: Solid St. Phys. {\bf 13} (1980) 3895.
\bibitem{}     Rastelli E and Lindg\.{a}rd P-A, 1979 J. Phys. C: Solid St.
              Phys. {\bf 12} (1979) 1899.
\bibitem{}     Sudha Gopalan and M. G. Cottam, Phys. Rev. {\bf B 42}
              No. 1  (1990) 624.
\bibitem{}    C. F. Lo, K. K. Pan and Y. L. Wang, J. Appl. Phys. {\bf 70}
                (10) (1991) 6080.
\bibitem{}    K. K. Pan  and Y. L. Wang, J. Appl. Phys.
              {\bf 73} (10) (1993) 6099.
\bibitem{}    Lei Zhou and Ruibao Tao, Phys. Letts. A
               {\bf 214} No. 3,4 (1996) 199.
\bibitem{}    T. Holstein and H. Primakoff, Phys. Rev. {\bf 59}
                (1940) 1098.
\bibitem{}    A. K. Battacharjee, B. Coqblin, R. Jullien, M. Plischke,
              D. Zobin and M. J. Zuckermann, J. Phys. F {\bf 8} (1978)
                1793.
\end{thebibliography}
\end{document}